\documentclass[aps,pra,twocolumn,a4paper,showpacs,superscriptaddress,floatfix,10pt]{revtex4-1}

\usepackage[pdftex]{graphicx}
\usepackage{dcolumn}
\usepackage{bm}
\usepackage[usenames, dvipsnames]{color}
\usepackage{comment}

\usepackage[T1]{fontenc}

\usepackage{amsfonts}
\usepackage{amssymb}
\usepackage{amsmath}
\usepackage{amsthm}
\usepackage{multirow}

\usepackage[scr=boondoxo, scrscaled=1.05]{mathalfa}

\usepackage{subfigure}

\usepackage{array}
\newcolumntype{L}[1]{>{\raggedright\let\newline\\\arraybackslash\hspace{0pt}}m{#1}}
\newcolumntype{C}[1]{>{\centering\let\newline\\\arraybackslash\hspace{0pt}}m{#1}}
\newcolumntype{R}[1]{>{\raggedleft\let\newline\\\arraybackslash\hspace{0pt}}m{#1}}
\usepackage{tabularx}

\def\KeyWord#1{$\backslash$\IfColor{$\!\!$\textRed{#1}\textBlack}{#1}$\!\!$}

\def\red#1{#1}

\usepackage{wasysym}

\newcommand{\be}{\begin{equation} }
\newcommand{\ee}{\end{equation} }
\newcommand{\ba}{\begin{eqnarray} }
\newcommand{\ea}{\end{eqnarray} }

\def\em{\it}

\newcommand{\bit}{\begin{itemize}}
\newcommand{\eit}{\end{itemize}}
\newcommand{\ben}{\begin{enumerate}}
\newcommand{\een}{\end{enumerate}}

\newcommand{\Bw}{{\vec{\omega}}}
\newcommand{\Bt}{{\vec{\theta}}}

\def\Im{\mathrm{Im}}

\def\PQ{P_\mathrm{Q}}

\def\bra#1{\langle#1|}
\def\ket#1{|#1\rangle}
\def\braket#1#2{\langle#1|#2\rangle}
\def\qexp#1#2{\bra{#2}#1\ket{#2}}
\def\cexp#1{\langle#1\rangle}

\begin{document}
\title{Half-integer quantized topological response in quasiperiodically driven quantum systems}

\author{P. J. D. Crowley}
\email{philip.jd.crowley@gmail.com}
\affiliation{Department of Physics, Boston University, Boston, MA 02215, USA}

\author{I. Martin}
\affiliation{Materials Science Division, Argonne National Laboratory, Argonne, Illinois 60439, USA}

\author{A. Chandran}
\affiliation{Department of Physics, Boston University, Boston, MA 02215, USA}

\date{\today}

\begin{abstract}
A spin strongly driven by two harmonic incommensurate drives can pump energy from one drive to the other at a quantized average rate, in close analogy with the quantum Hall effect. 
The pumping rate is a non-zero integer in the topological regime, while the trivial regime does not pump. 
The dynamical transition between the regimes is sharp in the zero-frequency limit and is characterized by a Dirac point in a synthetic band structure.
We show that the pumping rate is {\em half-integer} quantized at the transition and present universal Kibble-Zurek scaling functions for energy transfer processes.
Our results adapt ideas from quantum phase transitions, quantum information and topological band theory to non-equilibrium dynamics, and identify qubit experiments to observe the universal linear and non-linear response of a Dirac point in synthetic dimensions. 

\end{abstract}

\maketitle

\paragraph*{Introduction: } 
A wide variety of classical and quantum systems undergo second-order phase transitions in equilibrium~\cite{stanley1971phase,domb1972phase,binney1992theory,goldenfeld1992lectures,chaikin1995principles,sachdev2011quantum}.
Near such transitions, a universal coarse-grained description emerges; this predicts, for example, the same fluctuations for the fluid density near the liquid-gas critical point as the magnetization near the Ising paramagnet-ferromagnet transition.
Far from equilibrium, the description of phase transitions is more complicated.
Although a number of dynamical transitions have been observed in driven dissipative systems~\cite{wang1973phase,henkel2008non,eisert2010noise,baumann2010dicke,karl2013tuning,sieberer2013dynamical,carr2013nonequilibrium,tauber2014critical,marcuzzi2014universal,raftery2014observation,klinder2015dynamical,marino2016driven,tauber2017phase,casteels2017critical}, comparatively little is known about their general theory.

\begin{figure}
    \centering
    \includegraphics[width=0.96\columnwidth]{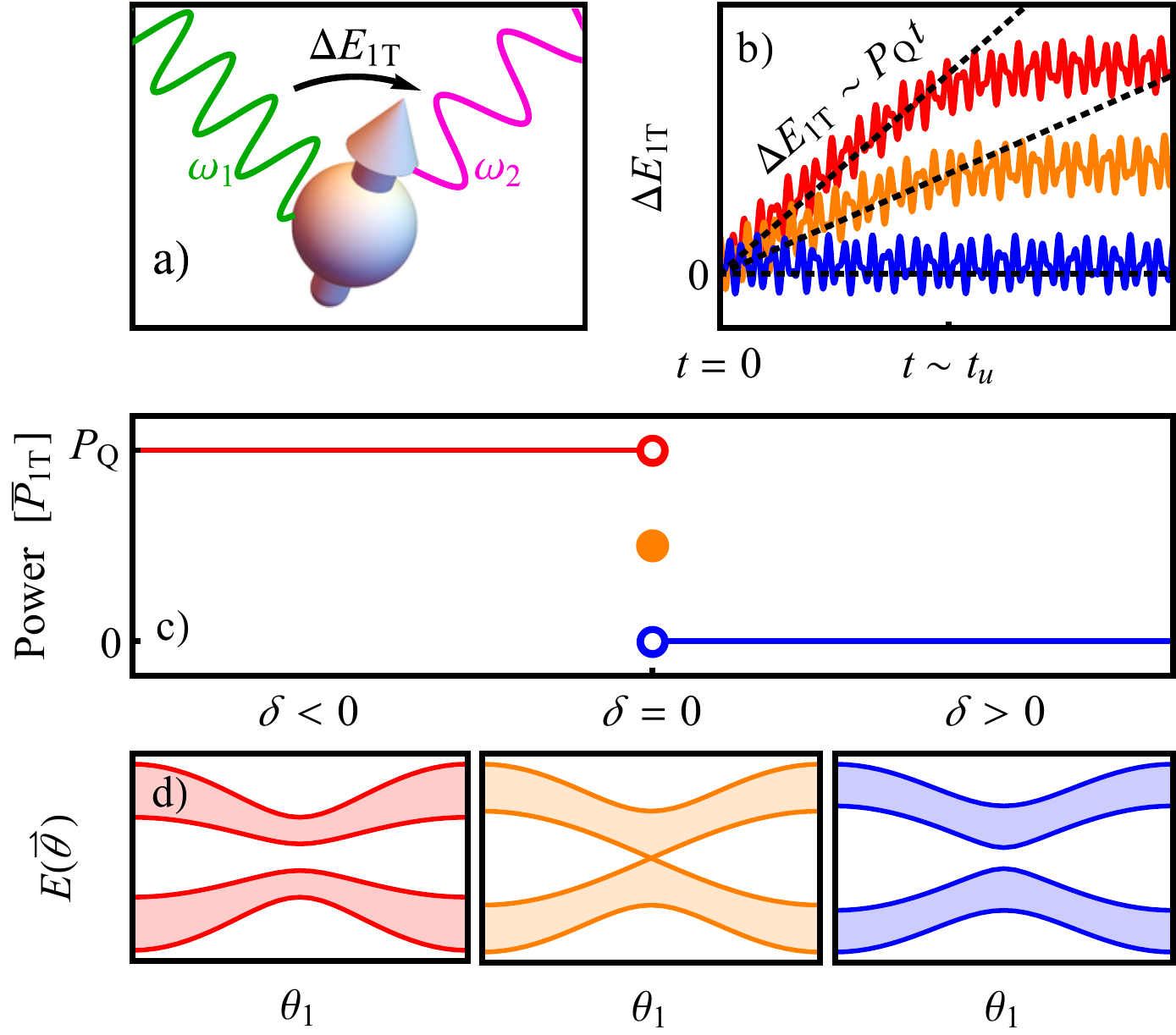}
    
    \caption{a) A spin driven by two incommensurate drives can transfer energy $\Delta E_{1\mathrm{T}}$ from one drive to the other.  b) The mean rate of energy transfer (power) is topologically quantized up to the unlock time $t_\mathrm{u}$. c) For $t \ll t_\mathrm{u}$, the mean power jumps from $P_\mathrm{Q} = \tfrac{\omega_1\omega_2}{2\pi}$ in the topologically nontrivial  regime ($\delta <0$) to zero in the trivial regime ($\delta>0$). At the transition ($\delta=0$), the mean power is $P_\mathrm{Q}/2$. d) The instantaneous energies of the Hamiltonian (Eq.~\eqref{Eq:HamBHZ}) organized into a band-structure as a function of $\theta_1$. At the transition, the band-structure contains a Dirac point which controls the universal low-frequency response.
    }
    \label{fig:Multipanel}
\end{figure}

Here we consider a dynamical phase transition in probably the simplest possible setting: a spin-$\tfrac12$ driven by two  drives with incommensurate frequencies. The drives may be produced by two optical cavities prepared in coherent states (Fig.~\ref{fig:Multipanel}a). The driving increases the richness of the problem by introducing two `synthetic dimensions', which correspond to the photon numbers in the two cavities~\cite{ozawa2019topological,peng2018topological,peng2018time,martin2017topological,crowley2019topological}. More precisely, the stationary states of the two-tone driven spin-$\tfrac12$ are given by the stationary states of a two-dimensional synthetic tight-binding model in the presence of an electric field which is equal to the vector of drive frequencies $(\omega_{1},\omega_{2})$~\cite{shirley1965solution,sambe1973steady,ho1983semiclassical,verdeny2016quasi,ozawa2019topological,peng2018topological,peng2018time,martin2017topological,crowley2019topological,nathan2019topological}.

For a range of parameters, the tight-binding model in synthetic dimensions exhibits the quantum Hall effect~\cite{bernevig2013topological,martin2017topological,nathan2019topological,crowley2019topological}. Ref.~\cite{martin2017topological} first identified the corresponding response in the driven qubit system: an average integer quantized energy current between the two drives in a direction set by the polarization of the qubit (Fig.~\ref{fig:Multipanel}(a-c)). Ref.~\cite{crowley2019topological} further identified other quantized responses in generic two-tone driven qudits and the integer topological invariant controlling these effects.

In this article, we show that the dynamical transition between the topological (pumping) and the trivial (non-pumping) regimes exhibits a half-integer quantized energy pumping rate (Fig.~\ref{fig:Multipanel}(b-c)). The transition is sharp in the zero-frequency limit. At the transition, the band structure of the synthetic model in the absence of the electric field has a Dirac point (Fig.~\ref{fig:Multipanel}(d)). The half-integer quantization follows from the integrated Berry curvature of one of the bands \emph{excluding the Dirac point}.   

To observe the quantized energy current the spin's evolution must be nearly adiabatic and `locked' in an instantaneous eigenstate. Away from the zero-frequency limit, the pumping is thus a pre-thermal effect. Using Kibble-Zurek (KZ) arguments~\cite{kibble1976topology,zurek1985cosmological,polkovnikov2005universal,zurek2005dynamics,dziarmaga2005dynamics,deng2009dynamical,dziarmaga2010dynamics,biroli2010kibble,de2011universal,polkovnikov2011universal,chandran2012kibble,kolodrubetz2012nonequilibrium,campo2014universality}, we show the time $t_\mathrm{u}$ when the spin unlocks to diverge as:
\begin{align}
    t_\mathrm{u} \sim \sqrt{B_0 /\omega^{3} }\label{Eq:KZTime},
\end{align}
where $B_0$ is the typical amplitude of the instantaneous field, and $\omega = \sqrt{\omega_1 \omega_2}$ is the typical frequency of the drives. On the time scales larger than $ t_\mathrm{u}$, the spin's direction is effectively decoupled from the external drives, leading to zero average pumping rate (Fig.~\ref{fig:Multipanel}(b)).

Near the transition, we show that the pump power is universal when measured in units of the diverging time-scale $t_\mathrm{u}$. The pump power has two universal contributions: one of topological origin that is quantized to either an integer or a half-integer as $ t/ t_\mathrm{u} \to 0$, and another non-quantized contribution due to spin excitation. Drawing intuition from the action of time-reversal on Hall insulators, we isolate the two contributions and their associated \emph{universal Kibble-Zurek scaling functions} using time evolution with the Hamiltonian and its complex conjugate.
Experimentally, complex conjugation corresponds to reversing the circular polarization of one drive.
Incommensurately driven few-level quantum systems thus offer a unique window into the universal properties of the topological phase transitions of band insulators.


\paragraph*{Model:}
For concreteness, we work with the same model as Refs.~\cite{martin2017topological,nathan2019topological,crowley2019topological} in which a spin-$\tfrac12$ is driven by two circularly polarized magnetic fields:
\begin{equation}
\begin{aligned}
H(\vec{\theta}_t) &= -\tfrac{1}{2}\vec{B}(\vec{\theta}_t) \cdot \vec{\sigma}\\
\vec{B}\big(\vec{\theta}_t) &= B_0 ( \sin \theta_{t1} ,  \sin \theta_{t2}, 2+\delta -\cos \theta_{t1}  -  \cos \theta_{t2}  \big) ,
\end{aligned}
\label{Eq:HamBHZ}
\end{equation}
where $\vec{\theta}_t=(\theta_{t1}, \theta_{t2})$ is the vector of drive phases, $\theta_{ti} = \omega_i t + \theta_{0i}$, and the ratio of the drive frequencies $\omega_2/\omega_1$ is an irrational number. We assume that the ratio of drive frequencies is order one, so that $\omega = \sqrt{\omega_1\omega_2}$ is the single frequency-scale on which $H$ varies. The  spin operator is a vector of Pauli matrices,  $\vec{\sigma} = (\sigma_x, \sigma_y,\sigma_z)$; and the spin is initialised in the instantaneous ground state.

\begin{figure}[t]
    \centering
    \includegraphics[width=0.86\columnwidth]{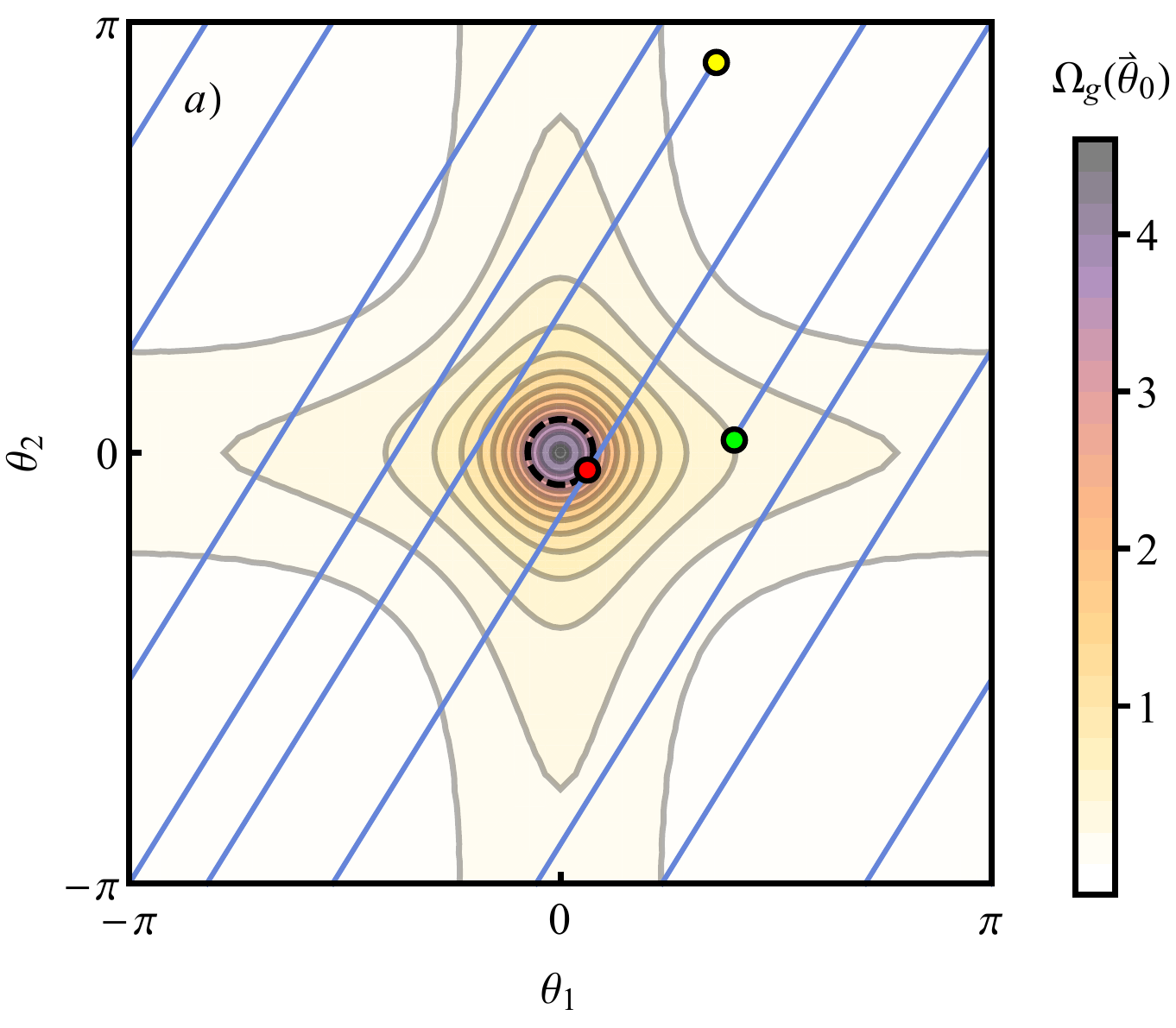}
    \caption{Contour plot of the Berry curvature of the instantaneous ground state band near the transition ($\delta = -1/3$). Blue lines depict the trajectory of the drive phases. The integrated Berry curvature sets the (average) energy pumping rate until the spin enters the excitation region $|\Bt_t|<\theta^*$ (dashed circle) for the first time (red point). }
    \label{fig:FZ}
\end{figure}

\emph{Instantaneous band structure:} At each $\Bt$, the instantaneous eigenstates of $H(\Bt)$ are either aligned or anti-aligned with the instantaneous magnetic field $\vec{B}(\Bt)$, with eigenvalues $\mp |\vec{B}(\Bt)|/2$ respectively. We organize these instantaneous eigenstates into a band structure on the torus $\vec{\theta}\in [0, 2\pi)^2$. The instantaneous band-structure describes the qubit dynamics for $t \ll t_\mathrm{u}$ when the spin's evolution remains adiabatic.

The Hamiltonian in Eq.~\eqref{Eq:HamBHZ} is engineered so that the instantaneous band-structure is topologically non-trivial.
Specifically, the instantaneous band-structure is identical to the momentum-space band-structure of a simple model of a quantum Hall insulator, the so-called half-BHZ model~\cite{qi2006topological,bernevig2006quantum,bernevig2013topological}. 
Consequently, the instantaneous ground state band (corresponding to the eigenvalue $-|\vec{B}(\Bt)|/2$) has a non-zero Chern number $C_g = 1$ for $-2<\delta<0$, and $C_g = 0$ for $\delta>0$. 

In the vicinity of the transition at $\delta=0$, there is a massive Dirac point in the band structure at $|\vec{\theta}|=0$ (Fig.~\ref{fig:Multipanel}(d)):
\begin{align}
    H(\vec{\theta}) =  - \tfrac{B_0}{2} (\theta_{1} \sigma_x + \theta_2 \sigma_y + \delta \sigma_z) + O(|\vec{\theta}|^2). \label{Eq:HamnearDirac}
\end{align}
Fig.~\ref{fig:FZ} is a density plot of the Berry curvature of the ground state band in the topological regime, close to the transition. 
The Berry curvature has two contributions: a piece that is smooth in $\delta$ and $\vec{\theta}$ and integrates to $\pi$; and a singular piece which concentrates into a delta function at $\Bt=\vec{0}$ and integrates to $-\textrm{sgn}(\delta) \pi$~\cite{bernevig2013topological,qi2006topological}.  

The drive phases follow trajectories of constant slope in $\vec{\theta}$-space (shown in blue in Fig.~\ref{fig:FZ}). Consider $t \ll t_\mathrm{u}$. Away from the Dirac point, the Berry curvature is small, and the spin state along the trajectory approximately follows the instantaneous ground state~\cite{weinberg2017adiabatic}. In the low frequency limit and with $\omega_2/\omega_1$ irrational, the trajectory uniformly samples this Berry curvature over time.

We show below (Eq.~\eqref{Eq:Pquantized}) that the average power is set by the integrated Berry curvature of the instantaneous ground state band \emph{before unlock}. Sufficiently far away from the transition, the trajectory thus samples the entire Berry curvature. At the transition, however, the spin unlocks before sampling the singular component associated with the Dirac point. Thus, the integrated Berry curvature that sets the average power is given by
\begin{align}
    C_g = \left\{ \begin{array}{cc}
    1 &\delta < 0 \\
    \tfrac12 & \delta = 0 \\
    0  &\delta > 0.
    \end{array}\right. 
    \label{Eq:Cgdelta}
\end{align}

\emph{Pump power:}
The instantaneous rate of energy transfer from drive 1 is given by~\cite{martin2017topological}
\begin{align}
    P_1 \equiv \omega_1 \cexp{\partial_{\theta_{1} } H },
\end{align}
with a corresponding expression for $P_2$. As the spin cannot absorb energy indefinitely, the net energy flux into the system time averages to zero, $[P_\mathrm{tot}]_t = [P_1]_t + [P_2]_t = 0$. Throughout, $[\cdot]_x$ denotes averaging with respect to variable $x$.

In the low frequency limit, $P_1$ is a sum of two terms, one analytic and one non-analytic in $\omega$. The analytic term is completely determined by the instantaneous values of $\vec{\theta}_t$, while the non-analytic terms depend on the entire history of the trajectory. As in the Landau-Zener problem, the analytic terms describe the perturbative ``dressing'' of the spin state over the instantaneous ground state~\cite{rigolin2008beyond,de2010adiabatic,weinberg2017adiabatic}. The non-analytic terms capture the non-adiabatic excitation processes between the dressed states. 
Below we refer to the leading order analytic term as $P_{1\mathrm{T}}$, as it is of topological origin, and non-analytic term as $P_{1\mathrm{E}}$, which is due to excitations.

\paragraph*{Topological contribution to pumping for $t \ll t_\mathrm{u}$:}
Let $|\tilde{g} (\Bt_t) \rangle$ be the spin state dressed to order $\omega$ above the instantaneous ground state. Thus,
\begin{equation}
\begin{aligned}
    i \frac{d |\tilde{g}  \rangle}{dt} &= H |\tilde{g} \rangle + O(\omega^2) \\
    \Rightarrow H |\tilde{g}\rangle &= i \omega_1 \ket{ \partial_{\theta_{1} } \tilde{g}} + i \omega_2 \ket{ \partial_{\theta_{2} }  \tilde{g}} + O(\omega^2). 
\end{aligned}
\label{Eq:gtildeSeq}
\end{equation}
where we have suppressed the time dependence of $\Bt$ for brevity. Using the product rule and Eq.~\eqref{Eq:gtildeSeq}, we obtain
\begin{equation}
\begin{aligned}
    P_{1\mathrm{T}} & = \omega_1 \qexp{\partial_{\theta_{1} } H }{\tilde{g}}  \\
    & =  \omega_1\left[\partial_{\theta_{1} } \qexp{ H }{\tilde{g}} - \bra{ \partial_{\theta_{1} } \tilde{g}}H\ket{\tilde{g}} - \bra{\tilde{g}}H\ket{ \partial_{\theta_{1} } \tilde{g}}\right] \\
    & = \omega_1 \partial_{\theta_{1} } \qexp{ H }{\tilde{g}} + \omega_1\omega_2 \Omega_{\tilde{g}}(\Bt)  ,
\end{aligned}
\label{eq:power}
\end{equation}
where $\Omega_{\tilde{g}}(\Bt) = 2 \Im \braket{ \partial_{\theta_{1} } \tilde{g}}{ \partial_{\theta_{2} } \tilde{g} }$ is the Berry curvature of the dressed spin state. 

The instantaneous power varies with the initial phase vector $\vec{\theta}_0$. 
Universal results about the spin dynamics at each $t$ follow upon initial phase averaging
\begin{equation}
\begin{aligned}
    \left.\right. [P_{1\mathrm{T}}]_{\vec{\theta_0}} &= [\partial_{\theta_{1t} } \qexp{ H }{\tilde{g}}]_{\vec{\theta_0}} + \omega_1\omega_2 [\Omega_{\tilde{g}}(\Bt_t)]_{\vec{\theta_0}}.
\end{aligned}
\label{Eq:P1TForm}
\end{equation}
The first term vanishes as it is a total derivative. The second term is the integrated Berry curvature of the dressed band prior to unlock (and thus excludes the Dirac point). As the integrated Berry curvature of the dressed and instantaneous bands are identical, we obtain:
\begin{align}
    [P_{1\mathrm{T}}]_{\vec{\theta_0}} = C_g \PQ \equiv C_g  \frac{ \omega_1 \omega_2}{2\pi} \label{Eq:Pquantized},
\end{align}
with $C_g$ given by Eq.~\eqref{Eq:Cgdelta}. \red{Formally Eq.~\eqref{Eq:Pquantized} holds at fixed $\delta$ as $\omega \to 0, t/t_\mathrm{u} \to 0$.
}

\paragraph*{Kibble-Zurek estimate for $t_\mathrm{u}$:}
The probability to transition to the dressed instantaneous excited state follows from the Landau-Zener result~\cite{zener1932non,landau1937theory,*landau1937theoryJ,*landau1937theory2,*landau1937theory2J,de2010adiabatic}
\begin{align}
     p_{\mathrm{exc}}  \sim \max_t \exp\left(-\frac{\pi|\vec{B}(\Bt_t)|^2}{|\partial_t \vec{B}|}\right).
     \label{Eq:LZResult}
\end{align}
Deep in the topological or trivial regimes, the spin's evolution thus remains adiabatic for an exponentially long time-scale $\sim  \exp\left(B_0/\omega\right)$.

Eq.~\eqref{Eq:LZResult} predicts that the spin unlocks from the field when the instantaneous gap squared becomes comparable or smaller than the rate of change of the field~\cite{kibble1976topology,zurek1985cosmological,polkovnikov2005universal,zurek2005dynamics,dziarmaga2005dynamics,dziarmaga2010dynamics,polkovnikov2011universal,chandran2012kibble,kolodrubetz2012nonequilibrium}. At the transition, the spin thus unlocks when
\begin{align}
    |\Bt_t| \lesssim \theta^* := \sqrt{\omega / B_0}.
    \label{Eq:thetastar}
\end{align} 
This relation defines the ``excitation region'' within the dashed circle in Fig.~\ref{fig:FZ}a. A typical spin trajectory enters the excitation region for the first time after $2\pi/\theta^*$ periods. We thus obtain the scaling of the unlock time $t_\mathrm{u} \sim (\omega \theta^*)^{-1} \sim \sqrt{ B_0 / \omega^{3} }$ previously stated in Eq.~\eqref{Eq:KZTime}. 


\paragraph*{Topological contribution to pumping for $t \gg t_\mathrm{u}$:}
At times much longer than the unlock time, non-adiabatic processes heat the spin.
In the initial phase ensemble, the populations in the (dressed) instantaneous ground and excited states thus become equal.
As the Chern numbers of the ground and excited state bands sum to zero, the ensemble averaged power $[P_{1\mathrm{T}}]_{\Bt_0} \to 0$ as $t/t_\mathrm{u} \to \infty$. 

\paragraph*{Excitation contribution to pumping:}
The non-adiabatic excitation of the spin results in a distinct contribution to the power $[P_{1\mathrm{E}}]_{\Bt_0}$. As the spin absorbs order $B_0$ energy from the drives over a time-scale $t_\mathrm{u}$
\begin{align}
    [P_{1\mathrm{E}}]_{\Bt_0} \sim B_0/t_\mathrm{u} \propto \omega^{3/2}, \quad t \lesssim t_\mathrm{u}. \label{Eq:P1EForm}
\end{align}
Unlike the topological contribution, the power due to excitation is non-analytic in $\omega$. The total pumped power is the sum of the topological and excitation contributions. 

A constant rate of excitation results in a linear increase of the excited state population in the initial phase ensemble at small $t/t_\mathrm{u}$. At late times, the populations become equal, and statistically the spin ceases to absorb energy from the drives. Thus, $[P_{1\mathrm{E}}]_{\Bt_0}\to 0$ as $t/t_\mathrm{u} \to \infty$. 

\paragraph*{Kibble-Zurek scaling functions:}
Within the Kibble-Zurek (KZ) scaling limit, the non-equilibrium dynamics of the spin becomes universal even beyond the unlock time. The KZ scaling limit involves taking $\omega, \delta \to 0$ together while measuring time in units of the diverging unlock time $t_\mathrm{u}$, and the drive frequency in units of the vanishing scale $\omega^*\sim B_0 \delta^2$~\cite{polkovnikov2005universal,deng2009dynamical,biroli2010kibble,de2011universal,polkovnikov2011universal,kolodrubetz2012nonequilibrium,chandran2012kibble}.
\red{The KZ scaling limit accesses the `critical fan' around the $\omega=0, \delta = 0$ transition, while Eq.~\eqref{Eq:Pquantized} applies deep within each `dynamical phase' (at fixed $\delta$) as $\omega \to 0$.}

In the KZ scaling limit, the radius of the excitation region $\theta^*$ becomes small and the Hamiltonian of the massive Dirac cone (Eq.~\eqref{Eq:HamnearDirac}) controls the excitation of the spin, and hence the decay of the topological component of the power.
The topological and excitation components of the power then take the following scaling forms
\begin{equation}
\begin{aligned}
    \left.\right.[P_{1\mathrm{E}}(t;\omega,\delta)]_{\Bt_0} &\sim \omega^{3/2} \mathscr{P}_{1\mathrm{E}}\left(t\, \omega^{3/2} \,; \delta\,\omega^{-1/2} \right),
    \\
    [P_{1\mathrm{T}}(t;\omega,\delta)]_{\Bt_0} &\sim \frac{\omega^{2}}{2 \pi} \mathscr{P}_{1\mathrm{T}}\left(t\, \omega^{3/2} \,; \delta\,\omega^{-1/2} \right).
    \label{eq:scalingFn}
\end{aligned}
\end{equation}
Above, $\mathscr{P}_{1\mathrm{E}}$ and $\mathscr{P}_{1\mathrm{T}}$ are scaling functions determined solely by the universality class of the transition in the instantaneous band structure. They capture the universal cross-over from the pre-thermal regime to the late-time infinite-temperature regime.

\begin{figure}[tb]
    \centering
    \includegraphics[width=0.95\columnwidth]{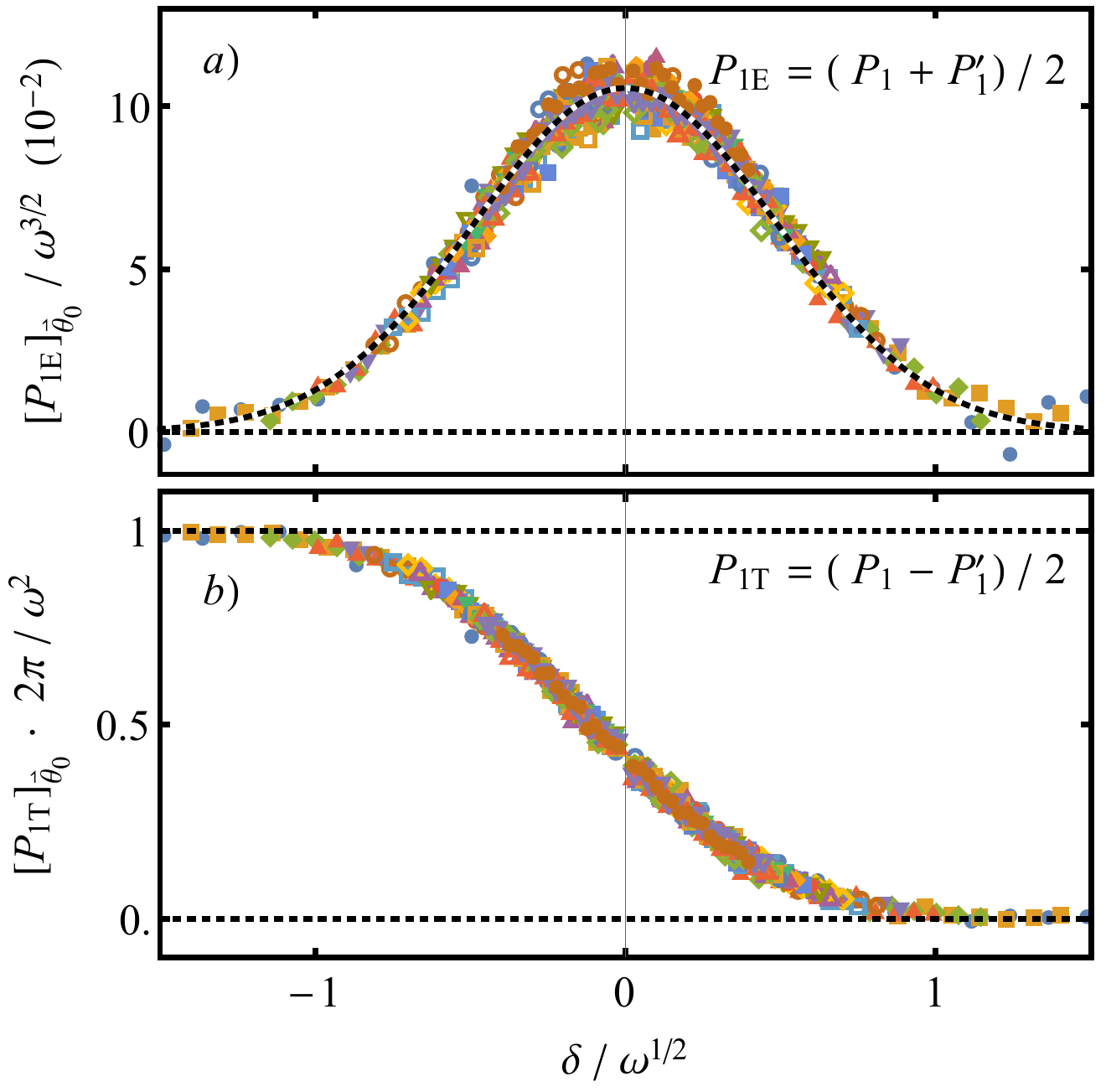}
    \caption{The phase and time averaged excitation (top) and topological (bottom) scaling functions of the power \red{averaged up to time} $t \omega^{3/2} = 1.72$ ($t/t_\mathrm{u} \approx 0.3$). As $p_{\mathrm{exc}} \propto \mathscr{P}_{1\mathrm{E}}$, the top panel is fit well by a Gaussian. The topological scaling function \red{in the bottom panel} decreases from one deep in the topological phase $\delta \ll - \omega^{1/2}$, to approximately one-half at the transition at $\delta=0$ and to zero in the trivial phase. \red{As $\delta$ is measured in units of $\omega^{1/2}$, the scaling function reproduces the step function in Eq.~\eqref{Eq:Pquantized} in the limit $\omega \to 0$ at fixed $\delta$.} Parameters: $B_0 = 1$, $\omega_2/\omega_1$ is the golden ratio and $20$ values of $ \omega \in [0.0025,0.1]$.}
    \label{fig:ScalingCollapse} 
\end{figure}

\paragraph*{Scaling functions for the Dirac transition:}
We now numerically extract the scaling forms $\mathscr{P}_{1\mathrm{E}}, \mathscr{P}_{1\mathrm{T}}$ for the Dirac transition in the half-BHZ model \red{by comparing the trajectories generated by $H(\Bt_t)$ and the complex-conjugated Hamiltonian $H^\prime(\Bt_t) = (H(\Bt_t))^*$.
Physically, $H^\prime$ is implemented by flipping the chirality of one of the circularly polarized drives. As this flips the sign of $[P_{1\mathrm{T}}]_{\Bt_0}$ alone, we can separate the topological and excitation contributions to the pumped power: }
\begin{equation}
\begin{aligned}
    {[P_{1\mathrm{E}}]_{\Bt_0}} &= \frac{1}{2}\left([P_{1}]_{\Bt_0} + [P_{1}^\prime]_{\Bt_0} \right) \\
    [P_{1\mathrm{T}}]_{\Bt_0} &= \frac{1}{2}\left([P_{1}]_{\Bt_0} - [P_{1}^\prime]_{\Bt_0} \right), \\
\end{aligned}
\label{eq:PEPT}
\end{equation}
Here $P_1^\prime = \omega_1\qexp{\partial_{\theta_1}H^\prime(\Bt_t)}{\psi_t^\prime}$ \red{is the instantaneous rate of energy transfer from drive 1 in the conjugated system} $i\partial_t\ket{\psi_t^\prime} = H^\prime(\Bt_t)\ket{\psi_t^\prime}$ when it is initialized in its instantaneous ground-state band $\ket{\psi_0^\prime} = (\ket{\psi_0})^*$. In the Supplemental Material, we obtain the same scaling functions for a different microscopic model with a Dirac transition, demonstrating universality.

In more detail, Eq.~\eqref{eq:PEPT} is obtained as follows. As the probability of excitation in Eq.~\eqref{Eq:LZResult} is invariant under complex conjugation, the ensemble populations of the instantaneous eigenstates are the same at each $t/t_\mathrm{u}$ for time evolution under $H$ and $H'$. Thus, $[P_{1\mathrm{E}}^\prime]_{\Bt_0} = [P_{1\mathrm{E}}]_{\Bt_0}$. The topological piece however changes sign, $[P_{1\mathrm{T}}^\prime]_{\Bt_0} = - [P_{1\mathrm{T}}]_{\Bt_0}$ as the Berry curvature changes sign under complex conjugation.

Fig.~\ref{fig:ScalingCollapse} shows the scaling functions across the transition for both contributions for small $t/t_\mathrm{u}$. The excitation scaling function is proportional to excitation probability for $t \ll t_\mathrm{u}$. As $p_{\mathrm{exc}} \sim \exp(-\delta^2 B_0/\omega)$, Fig.~\ref{fig:ScalingCollapse}(a) is thus well fit by a Gaussian centred at the transition (black-white dashed line).

Fig.~\ref{fig:ScalingCollapse}(b) shows $[P_{1\mathrm{T}}]/\PQ$ to obey a single-parameter scaling function at fixed $t/t_\mathrm{u}$, with the $[P_{1\mathrm{T}}]/\PQ \to 1(0)$ as $\delta \to -\infty(\infty)$ (Eq.~\eqref{Eq:Cgdelta}).
The intermediate value at $\delta = 0$ is close to $\tfrac12$, and becomes exactly $\tfrac12$ as $t/t_\mathrm{u} \to 0$.
For $\delta\omega^{-1/2} = \mathrm{O}(1)$, the component of the Berry curvature that is singular at the transition has a width comparable to that of the excitation region $\theta^*$.
Consequently, this component is partially sampled by spin trajectories before unlock and leads to the smooth universal function observed in Fig.~\ref{fig:ScalingCollapse}(b).
Note that Eq.~\eqref{Eq:Pquantized} is contained within the scaling function in Eq.~\eqref{eq:scalingFn} on taking $\omega \to 0$ at fixed $\delta$.

A technical comment: the data in Fig.~\ref{fig:ScalingCollapse} is time averaged over $t\omega^{3/2} \in [0,1.72]$ (in addition $\Bt_0$ averaging). \red{This reduces fluctuations due to the finite sampling of $\Bt_0$ and quasiperiodic micro-motion}, and slightly modifies the value of the scaling function near $\delta=0$. \red{Scaling functions without time averaging are discussed in the Supplemental Material.}

\begin{figure}
    \centering
    \includegraphics[width=0.95\columnwidth]{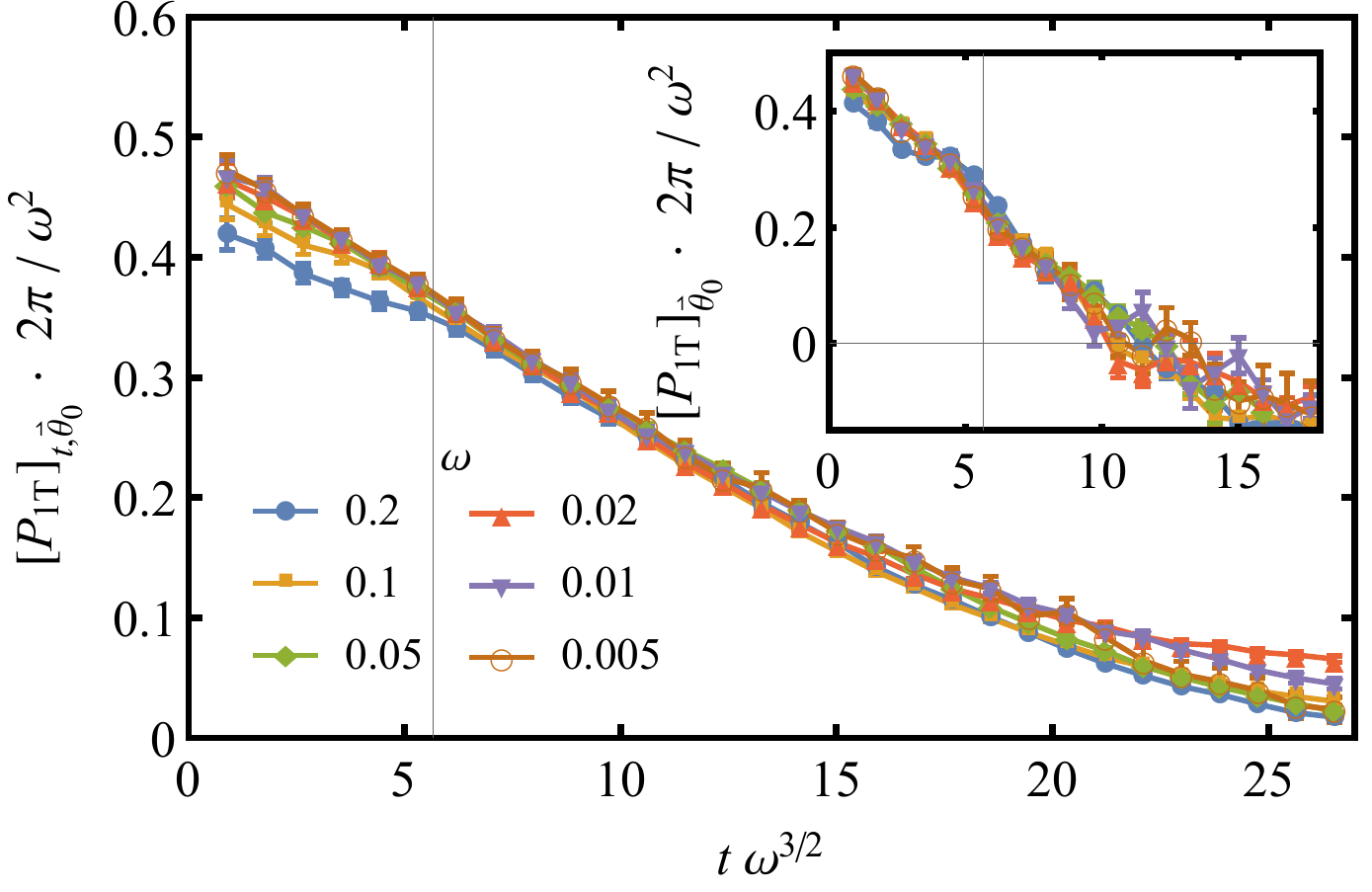}
    \caption{The topological scaling function at the transition with (main) and without time averaging (inset). Both plots show that the scaling function approaches $1/2$ as $t \omega^{3/2} \to 0$, and zero as $t \omega^{3/2} \to \infty$. The vertical line is a proxy for $t_\mathrm{u}$ (definition in text). \red{See the supplemental material for the excitation scaling function at the transition.}
    }
    \label{fig:HalfIntegerScaling}
\end{figure}

We now turn to the dependence of the scaling functions on $t/t_\mathrm{u}$. Fig.~\ref{fig:HalfIntegerScaling} (inset) shows $\mathscr{P}_{1\mathrm{T}} \sim [P_{1\mathrm{T}}]_{\Bt_0}/\PQ$  at the transition, whilst the main figure shows the power with additional averaging over the time interval $[0,t]$. The grey vertical line is a proxy for $t_\mathrm{u}$ and identifies the mean time up to which $|\partial_t \vec{B}|< |\vec{B}|^2$. 
\red{
Both plots show that, as $t/t_\mathrm{u} \to 0$, $\mathscr{P}_{1\mathrm{T}}$ approaches the quantized value of
\begin{equation}
    \mathscr{P}_{1\mathrm{T}}(0;0) = \tfrac{1}{2}\left( \left.C_g\right|_{\delta>0} + \left.C_g\right|_{\delta<0} \right) = \left.C_g\right|_{\delta=0} = \tfrac12
    \label{eq:quant}
\end{equation}}
Next, for $0 < t/t_\mathrm{u} \ll 1$, $\mathscr{P}_{1\mathrm{T}}$ linearly decreases. 
This linear decrease follows from the constant excitation rate in Eq.~\eqref{Eq:P1EForm} and the opposite sign of the pumping by the excited population. 
Finally, despite the negative turn at $ t\omega^{3/2} \approx 10$ in the inset, we find that $\mathscr{P}_{1\mathrm{T}}$ approaches zero for $t/t_\mathrm{u} \gg 1$, consistent with the main figure.

\paragraph*{Discussion:}
We have demonstrated that a simple quasiperiodically driven spin-$\tfrac12$ exhibits universal scaling behaviours characteristic of extended classical or quantum {\em equilibrium} systems in the vicinity of continuous phase transitions~\cite{cardy1996scaling,sachdev1999quantum}. 
Our results serve as a new example of the tantalising correspondence emerging between equilibrium systems and systems subject to periodic or quasiperiodic driving~\cite{ho1983semiclassical,luck1988response,jauslin1991spectral,blekher1992floquet,jauslin1992generalized,kitagawa2011transport,lindner2011floquet,jiang2011majorana,cayssol2013floquet,katan2013modulated,rudner2013anomalous,iadecola2013materials,delplace2013merging,kundu2014effective,grushin2014floquet,lababidi2014counter,chandran2016interaction,verdeny2016quasi,khemani2016phase,von2016phase,von2016phase2,roy2016abelian,else2016classification,nathan2017quantized,roy2017periodic,moessner2017equilibration,martin2017topological,baum2018setting,peng2018topological,mondragon2018quantized,kolodrubetz2018topological,dumitrescu2018logarithmically,peng2018time,nathan2019topological,peng2019floquet,bauer2019topologically,ozawa2019topological,hu2019dynamical,crowley2019topological,oka2019floquet}.

The KZ scaling functions also provide the universal non-linear response of a clean Dirac material in an electric field~\cite{green2005nonlinear}. Fourier transforming the $\vec{\theta}$ coordinates maps the model in Eq.~\eqref{Eq:HamBHZ} on to the real-space model of a Hall insulator (the half-BHZ model) with an additional electric field $(\omega_1, \omega_2)$. In this transformation, $\omega$ maps on to the magnitude of the electric field, the topological component of the power maps on to the Hall current, and the excitation component measures the population in the excited band due to dielectric breakdown when the insulator is initially at zero temperature.

Driven few-level systems can access other topological phase transitions in static systems using different driving protocols. 
Moreover, the KZ scaling theory can be  extended to include the effects of dissipation~\cite{nathan2019topological}, or counter-diabatic driving~\cite{crowley2019topological}.
Both effects increase the unlock time $t_\mathrm{u}$, and may simplify experimental access to the half-quantized response in solid-state and quantum optical platforms that host qubits~\cite{jelezko2004observation,buluta2011natural,clarke2008superconducting,dobrovitski2013quantum,kloeffel2013prospects,haffner2008quantum,devoret2004superconducting,langer2005long,taylor2005fault,trauzettel2007spin,gali2009theory,blatt2012quantum,harty2014high,wendin2017quantum,wang2017single}.

\paragraph*{Acknowledgements: } We are grateful to E.~Boyers, W.W.~Ho, C.~R.~Laumann, D.~Long, A.~Polkovnikov, D.~Sels and A.~Sushkov for useful discussions. This work was supported by NSF DMR-1752759 (A.C. and P.C.), and completed at the Aspen Center for Physics, which is supported by the NSF grant PHY-1607611. A.C. and P.C. acknowledge support from the Sloan Foundation through the Sloan Research Fellowship. Work at Argonne was supported by the Department of Energy, Office of Science, Office of Basic Energy Sciences, Materials Science and Engineering Division.

\bibliography{QPD_bib}

\clearpage

\section*{Supplemental Material}

\subsection*{Universality of scaling functions at the Dirac transition}
\label{app:universality}

\begin{figure}[t!]
    \centering
    \includegraphics[width=0.95\columnwidth]{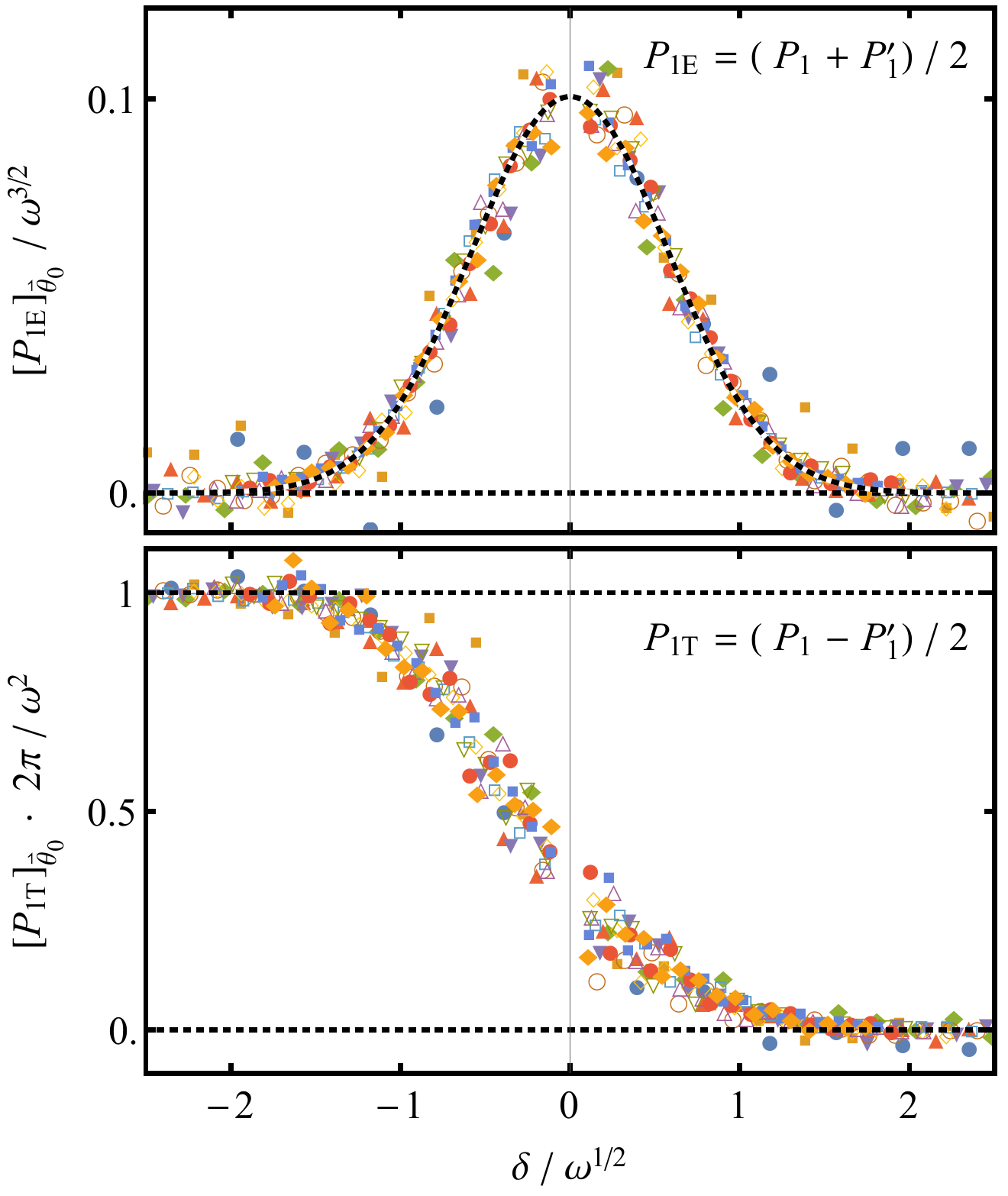}
    \caption{
    The phase averaged excitation (top) and topological (bottom) scaling functions of the power at $t \omega^{3/2} = 3.43$. In re-scaled units, the scaling functions are identical to those in Fig.~\ref{fig:ScalingCollapse}. Parameters: $\vec{B}$ given by~\eqref{eq:supp_model} with $(n_1,n_2) = (1,3)$ and $\phi = 4\pi/5$, $B_0 = 1$, $\omega_2/\omega_1$ is the golden ratio and we use $13$ values of $ \omega \in [1.6 \times 10^{-4}, 3 \times 10^{-2}]$.
    }
    \label{fig:SuppScalingCollapse}

\end{figure}

The Hamiltonian used in the main text (Eq.~\eqref{Eq:HamBHZ}) is a well-known model~\cite{qi2006topological,hasan2010colloquium,bernevig2013topological,bansil2016colloquium}, which has several special symmetries. Universality requires that the scaling functions $\mathscr{P}_{1\mathrm{E}}, \mathscr{P}_{1\mathrm{T}}$ do not depend on these symmetries. Here we repeat the numerical analysis shown in the main text using a model which lacks these symmetries. We obtain scaling collapse in the KZ scaling limit and identical scaling functions to that in Fig.~\ref{fig:SuppScalingCollapse}.

Symmetries of the Hamiltonian~\eqref{Eq:HamBHZ} relate non-trivial actions in $\Bt$-space to rotations of the spin:
\begin{equation}
\begin{aligned}
    H(-\Bt) &= \sigma_z H(\Bt) \sigma_z \\
    -(H(\Bt))^* &= \sigma_y H(\Bt) \sigma_y \\
    H(\theta_{2},-\theta_{1}) & = U_{\pi/2} H(\theta_{1},\theta_{2}) U_{\pi/2}^{\dag}.
\end{aligned}
\label{eq:special}
\end{equation}
Here $U_\phi = \exp\left(-i \phi \sigma_z/2\right)$. Furthermore, in the vicinity of the Dirac point at $\delta = 0$,
\begin{equation}
    H(\Bt) = 
    -\frac{B_0}{2}
    \begin{pmatrix}
    \theta_1 \\
    \theta_2 \\
    0
    \end{pmatrix}
    \cdot \vec\sigma
    -
    \frac{B_0}{4}
    \begin{pmatrix}
    0 \\
    0 \\
     |\Bt|^2
    \end{pmatrix}
    \cdot \vec\sigma
    + 
    O(|\Bt_t|^3),
\end{equation}
and the final relation in Eq.~\eqref{eq:special} holds for any rotation angle
\begin{equation}
    H(R_\phi \Bt_t) = U_\phi H(\Bt_t) U_\phi^\dag +  O(|\Bt_t|^3).
\end{equation}
Above $R_\phi$ is a rotation by angle $\phi$ in $\Bt$-space
\begin{equation}
     R_\phi = \begin{pmatrix}\cos \phi & -\sin \phi\\ \sin \phi & \cos \phi \end{pmatrix}.
\end{equation}

To break the symmetries in Eq.~\eqref{eq:special}, consider a spin-$\tfrac12$ driven by the magnetic field $\vec{B}(\Bt_t)$
\begin{equation}
    \vec{B}(\Bt) = 
   B_0 \begin{pmatrix}
    \sin(n_1\theta_{1}-\phi) + \sin\phi \\ 
    \sin(n_2\theta_{2}-\phi) + \sin\phi \\
    2 + \delta - \cos \theta_{1} - \cos \theta_{2}
    \end{pmatrix}.
    \label{eq:supp_model}
\end{equation}
For $\cos\phi \neq 0$, $n_1 \neq n_2$ there is an asymmetric Dirac point at $\delta=0, \Bt=\vec{0}$,
\begin{align}
    H(\Bt) = -\frac{B_0 \cos \phi}{2}
    \begin{pmatrix}
    n_1 \theta_1  \\
    n_2 \theta_2  \\
    0
    \end{pmatrix}
    \cdot \vec\sigma
    &-
    \frac{B_0}{4}
    \begin{pmatrix}
      n_1^2 \theta_1^2 \sin \phi  \\
      n_2^2 \theta_2^2 \sin \phi \\
     |\Bt|^2
    \end{pmatrix}
    \cdot \vec\sigma \nonumber \\
    &+ O(|\Bt|^3).
\end{align}
Fig~\ref{fig:SuppScalingCollapse} shows the scaling collapse obtained for the topological and trivial contributions to the power. In rescaled units, the scaling functions are identical to those obtained in the main text (Fig~\ref{fig:ScalingCollapse}). 

\red{
\subsection*{Additional scaling data without time averaging}
}
\begin{figure}
    \centering
        \includegraphics[width=0.95\columnwidth]{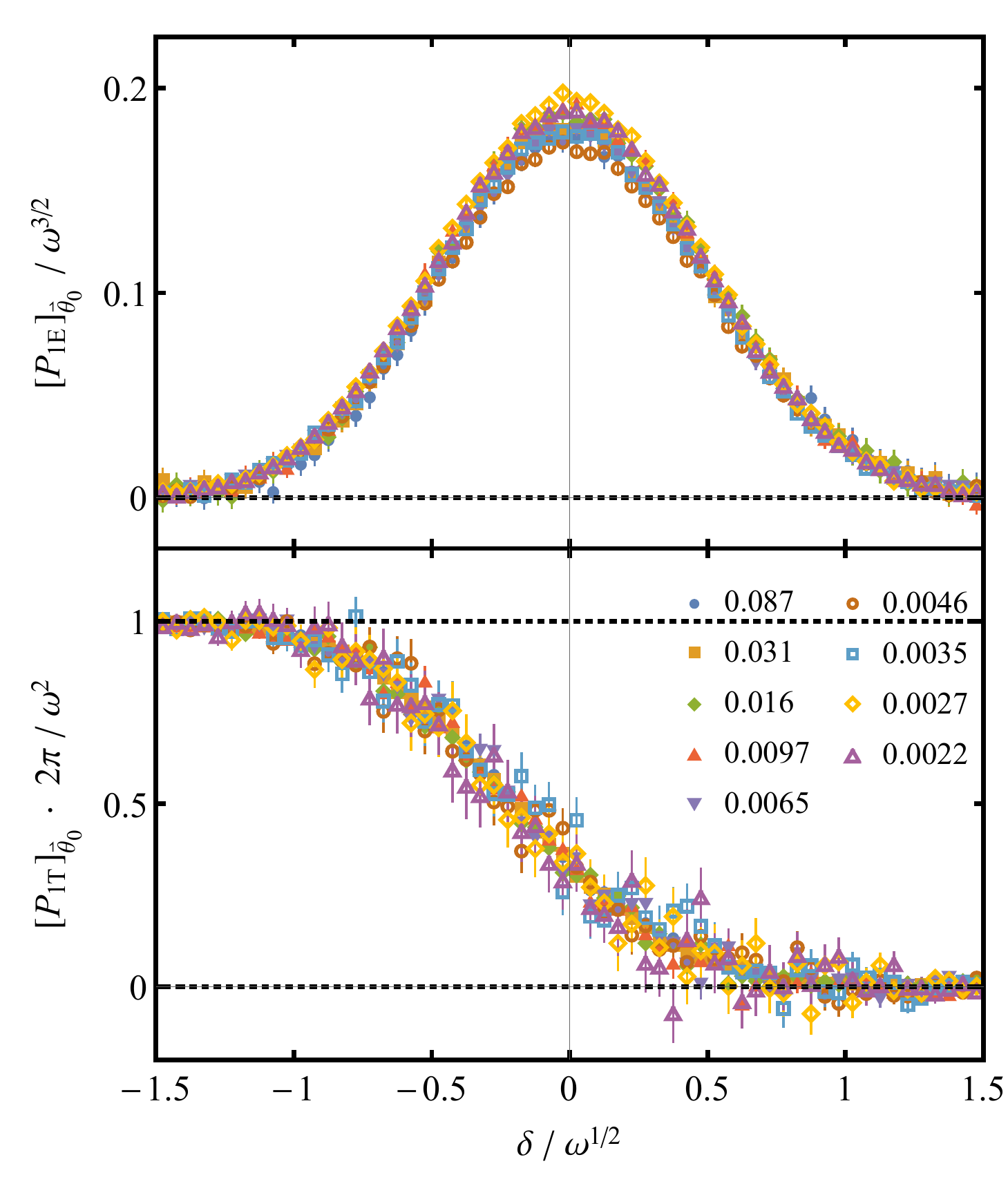}
    \caption{
    \red{
    The phase averaged excitation (top) and topological (bottom) scaling functions of the power at $t \omega^{3/2} = 3.53$ ($t/t_\mathrm{u} \approx 0.3$). The topological scaling function decreases from one deep in the topological phase $\delta \ll - \omega^{1/2}$, to approximately one-half at the transition at $\delta=0$ and on to zero in the trivial phase $\delta \gg - \omega^{1/2}$. The scaling functions are found to be qualitatively similar to those of Fig.~\ref{fig:ScalingCollapse} where the addition of time averaging leads to a reduction in noise due to the finite $\Bt_0$ sampling. Error bars indicate the standard error of the mean. Parameters: $B_0 = 1$, $\omega_2/\omega_1$ is the golden ratio and $9$ values of $ \omega \in [0.0022,0.087]$ (legend inset). 
    }}
    \label{fig:SuppScalingCollapseUnaveraged}

\end{figure}
\red{
In the main text, we presented scaling collapse for the topological ($[P_{1\mathrm{T}}]_{\Bt_0,t}$) and excitation  ($[P_{1\mathrm{E}}]_{\Bt_0,t}$) components of the pump power after initial phase averaging \emph{and} time averaging. Although the initial phase averaging is essential to define universal dynamical response functions, the time averaging is not (c.f. Eq.~\eqref{eq:scalingFn}). Instead, we performed time averaging to reduce the fluctuations of the pump power due to the finite sampling of $\Bt_0$ and quasiperiodic micro-motion. For completeness, we present the scaling functions in Fig.~\ref{fig:ScalingCollapse} and Fig.~\ref{fig:HalfIntegerScaling} without additional time averaging below. 
}

\red{
Fig.~\ref{fig:SuppScalingCollapseUnaveraged} presents the excitation (top) and topological (bottom) scaling functions at fixed $t/t_u\approx 0.6$ (or $t = 3.53 / \omega^{3/2}$). The different series correspond to nine different values of frequency $0.087 \leq \omega \leq 0.0022$ that were obtained by time evolution with $H$ and $H'$ (see Eq.~\eqref{eq:PEPT}). The scaling functions in Fig.~\ref{fig:SuppScalingCollapseUnaveraged} are qualitatively similar to those in Fig.~\ref{fig:ScalingCollapse}.
}

\red{In more detail, the $[P_{1\mathrm{T}}]_{\Bt_0}$ values are found to be noisier than $[P_{1\mathrm{E}}]_{\Bt_0}$, as they are sub-leading to $[P_{1\mathrm{E}}]_{\Bt_0}$ term, and any errors in the deducting off the value of $[P_{1\mathrm{E}}]_{\Bt_0}$ are thus higher order in $\omega$ than $[P_{1\mathrm{T}}]_{\Bt_0}$, and so must be made small through $\Bt_0$ averaging. 
}

\red{
As a technical comment, we choose the values of $\omega$ in Fig.~\ref{fig:SuppScalingCollapseUnaveraged} so that $\omega_1 t \in 2 \pi (\mathbb{N} + \tfrac12)$. We find that choosing $t$ such that $\omega_1 t \mod 2 \pi$ is equal for each series of data leads to a reduction in residual fluctuations in $[P_{1\mathrm{E}}]_{\Bt_0}$. However, this choice of $\omega_1 t \in 2 \pi (\mathbb{N} + \tfrac12)$ is inconsequential for $[P_{1\mathrm{T}}]_{\Bt_0}$, as we find the same quality of collapse of $[P_{1\mathrm{T}}]_{\Bt_0}$ as is found in Fig.~\ref{fig:SuppScalingCollapseUnaveraged} without this additional requirement.
}

\begin{figure}
    \centering
        \includegraphics[width=0.95\columnwidth]{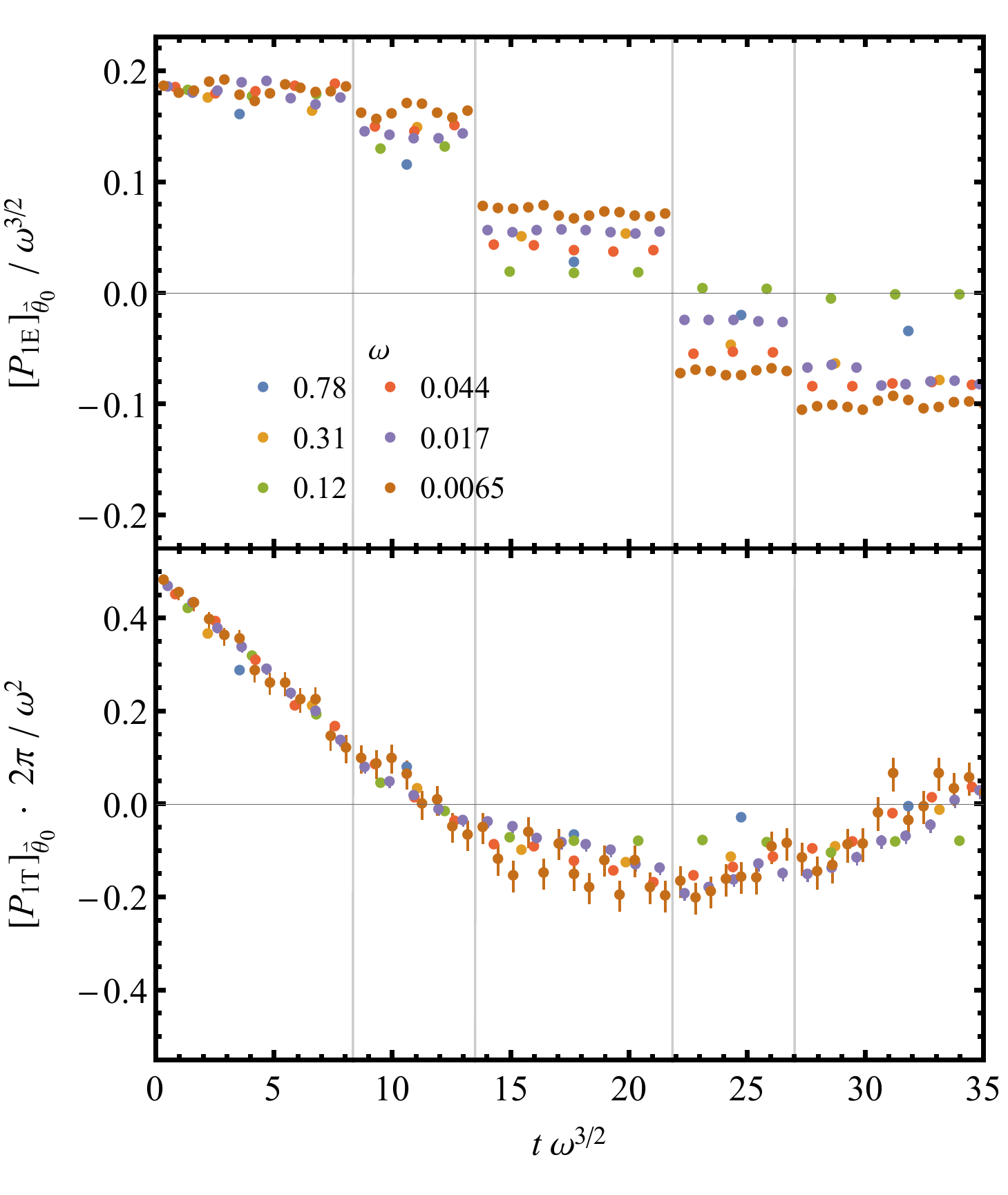}
    \caption{
    \red{
    The phase averaged excitation (top) and topological (bottom) scaling function at the transition. The topological component can be seen to approach $[P_{1\mathrm{T}}]_{\Bt_0} \cdot 2 \pi / \omega^2 = 1/2$ as $t \omega^{3/2} \to 0$. The vertical lines correspond to those obtained from the scaling collapse in Fig.~\ref{fig:SuppSteps}. As before $B_0 = 1$, $\omega_2/\omega_1$ is the golden ratio and $\omega$ values are shown in legend (inset), error bars indicate the standard error of the mean due to $\Bt_0$ averaging.
    }
    }
    \label{fig:SuppScalingCollapseUnaveraged2}

\end{figure}

\begin{figure}
    \centering
        \includegraphics[width=0.95\columnwidth]{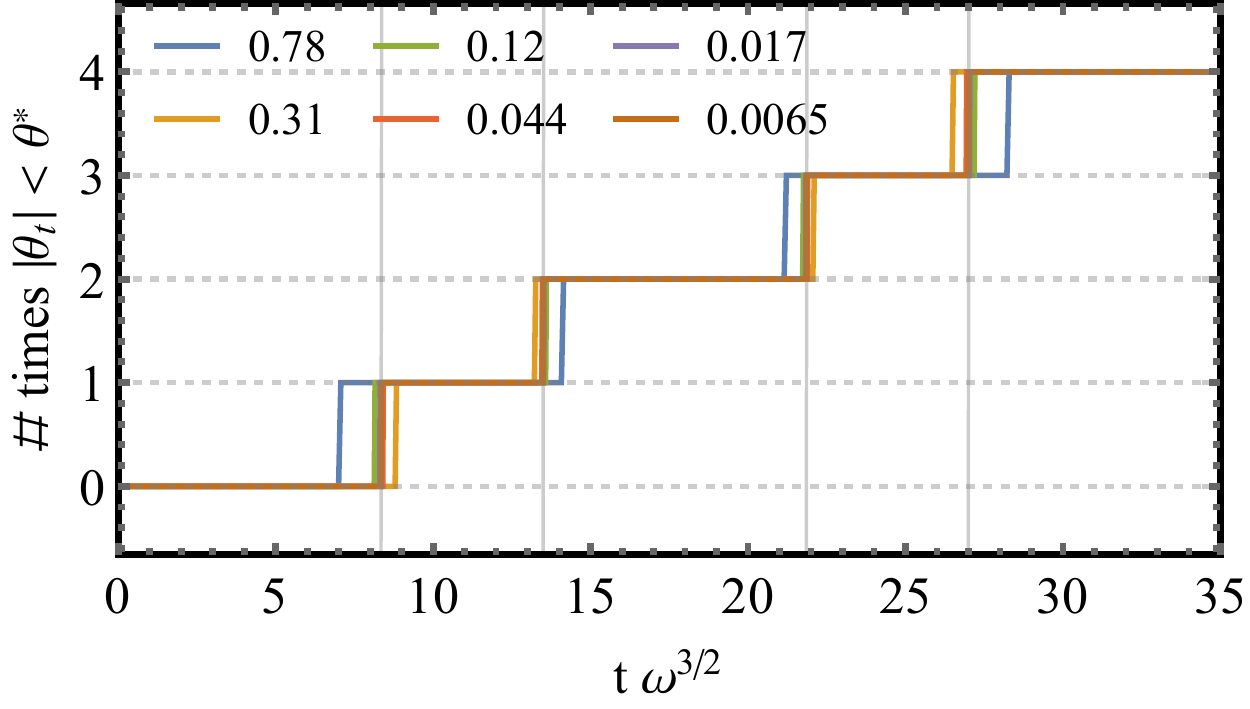}
    \caption{
    \red{
    \emph{Number of previous visits to the excitation region.} At a time $t$, from the ensemble of initial conditions we select the state which is at the Dirac point $\Bt_t = \vec{0}$. For these initial conditions we then ask how many times the state has previously visited the excitation region $|\Bt_s|<\theta^*$ for $0 < s < t$. Under discrete rescaling the resulting function collapses, and has step discontinuities which map onto the jumps in Fig.~\ref{fig:SuppScalingCollapseUnaveraged2}. For this plot we use $\theta^* = 3.5 \sqrt{B_0/\omega}.$ $\omega$ values (inset) are the same as in Fig.~\ref{fig:SuppScalingCollapseUnaveraged2}.
    }
    }
    \label{fig:SuppSteps}

\end{figure}

\red{
Fig.~\ref{fig:SuppScalingCollapseUnaveraged2} shows the scaling function at the transition without time averaging for the excitation component $[P_{1\mathrm{E}}]_{\Bt_0}$ (upper plot) and the topological component $[P_{1\mathrm{T}}]_{\Bt_0}$ (lower plot). The top panel exhibits approximate scaling collapse at accessible integration times, note that within each ``step'' of the plot, the trend is monotonic in $\omega$, providing indication of collapse in the limit of $\omega \to 0$. This scaling collapse is found to be discrete, and we have used frequencies $\omega \propto \tau^{2n}$ where $\tau = \tfrac12 (1+\sqrt5)$ is the golden ratio. The necessity for such discrete re-scaling, indicative of fractality, is a typical feature of the scaling theory of quasiperiodic systems (a most famous example is found in Hofstadter's butterfly~\cite{hofstadter1976energy}, the spectrum of the ``Almost Mathieu operator''). As before data is plotted so that $\omega_1 t \in 2 \pi (\mathbb{N} + \tfrac12)$ to remove residual fluctuations.
}

\red{
The scaling function for $[P_{1\mathrm{E}}]_{\Bt_0}$ exhibits remarkable step-like features. Their origin is as follows. At $t=0$, we prepare an ensemble of $N$ spins in their instantaneous ground states at $N$ values of the initial phase vector that are uniformly distributed on the initial phase torus. Let us assume that the excitation rate is zero outside the excitation region $|\Bt_t | < \theta^\star$. Then, in each time interval $\delta t$ for $ t\ll t_u$,  a number $N_\textrm{exc} \approx N \cdot (\theta^\star/ 2 \pi) \cdot (\omega \delta t/ 2 \pi)$ of spins enter the excitation region. As each such spin absorbs some constant energy from each drive, the ensemble averaged power $[P_{1\mathrm{E}}]_{\Bt_0}$ is approximately constant in time for $ t \ll t_u$. On a timescale $t \approx t_u$ this situation discretely changes. After a time $t \sim t_u$ the spins entering the excitation region will have already previously entered the excitation region, and are already excited. Already excited spins absorb a different amount of energy from each of the drives as they traverse the excitation region, and thus make a different contribution to $[P_{1\mathrm{E}}]_{\Bt_0}$. Thus, when this happens $[P_{1\mathrm{E}}]_{\Bt_0}$ exhibits a discrete step. Similar discrete events follow at later times, separated by $t \sim t_u \sim \omega^{3/2}$. This explains the step-like features in the excitation component of the scaling function of the pump power at discrete values of $t/t_u$ in the top panel of Fig.~\ref{fig:SuppScalingCollapseUnaveraged2}. 
}

\red{
We can verify our explanation of the steps in Fig.~\ref{fig:SuppScalingCollapseUnaveraged2}. To calculate the expected position of the steps, at each time $t$ we can select the spin in the ensemble which is at the Dirac point $\Bt_t = 0$. This simply fixes $\Bt_0 = (-\Bw t) \mod 2 \pi$. We can the calculate the number of unique visits it has previously made to the excitation region. This is given by the number of unique solutions $t_n$ to the equation $|\Bt_{t_n}|< \theta^*$ for $0<t_n<t$ where two solutions $t_n, t_m$ are equivalent if $\Bt_{s}$ does not leave the excitation region for $s \in [t_n,t_m]$. The number of previous unique visits is plotted in Fig.~\ref{fig:SuppSteps}, the steps in this function occur at the times $t_n$, and correspond closely with the steps in Fig.~\ref{fig:SuppScalingCollapseUnaveraged2} (vertical bars in both plots).
In these plots we have used an arbitrary definition of $\theta^* = 3.5 \sqrt{B_0/\omega}$ which successfully predicts the positions of the most significant step features in Fig.~\ref{fig:SuppScalingCollapseUnaveraged2}, and thus is sufficient to illustrate their origin. A more refined analysis would also consider precisely how close to the Dirac point each previous visit to the excitation region was rather than using an arbitrary cut-off as here. Such analysis would provide information on the step heights as well.
}

\red{
The discrete events have a less significant effect on the scaling function of the topological component of the pump power. ${[P_{1\mathrm{T}}]_{\Bt_0}}$ behaves qualitatively similarly to its time averaged equivalent in the main text, see Fig.~\ref{fig:HalfIntegerScaling}. The only notable new feature is the overshoot of ${[P_{1\mathrm{T}}]_{\Bt_0}}$ into negative values on a timescale $t \sim t_\mathrm{u}$, before converging to ${[P_{1\mathrm{T}}]_{\Bt_0}} \to 0$ at late times. We discuss this feature. At $t = 0$ the system is initialised in the ground state band, and thus initially pumps with power ${[P_{1\mathrm{T}}]_{\Bt_0}} = \tfrac12$. At early times excitation events transfer the population of the ground state band into the excited band. As this process is not random it can result in an excess of population in the excited band, which pumps in the opposite direction and hence ${[P_{1\mathrm{T}}]_{\Bt_0}} < 0$. However, the quasi-periodic exploration of the $\Bt$-torus is not sufficiently structured to prevent ${[P_{1\mathrm{T}}]_{\Bt_0}} \to 0$ at late times, as expected.
}


\end{document}